\documentstyle[12pt]{article}

\global\arraycolsep=1pt
\begin{document}
\baselineskip 18pt
\begin{titlepage}
\hfill  hep-th/9704004

\hfill  IFUM 527/FT

\hfill  April, 97

\begin{center}
\hfill
\vskip .4in
{\large\bf Infrared Stability of $N=4$ Super Yang-Mills Theory\footnotemark}
\footnotetext{Partially supported by EEC, Science Project ERBFMRX-ct96-0045.}
\end{center}
\vskip .4in
\begin{center}
{\large Michela Petrini\footnotemark}
\footnotetext{email michela.petrini@mi.infn.it}

\vskip .1in
{\em Dipartimento di Fisica, Universit\`a di Milano and Istituto Nazionale
di Fisica Nucleare (INFN) - Sezione di Milano, via Celoria 16,
I-20133 Milano, Italy}

\end{center}

\begin{center}
{\bf Abstract}
\end{center}
\begin{quotation}
\noindent We study the infrared perturbative properties of a class of
non supersymmetric gauge theories
with the same field content of $N=4$ Super Yang-Mills and we show that
the $N=4$ supersymmetric model represents an IR unstable fixed point for
the renormalization group equations.
\end{quotation}
\vspace{4pt}
\end{titlepage}
\noindent
\vfill
\eject

\newcommand{\be}{\begin{equation}}
\newcommand{\ee}{\end{equation}}
\newcommand{\ba}{\begin{eqnarray}}
\newcommand{\ea}{\end{eqnarray}}
\newcommand{\ban}{\begin{eqnarray*}}
\newcommand{\ean}{\end{eqnarray*}}
\newcommand{\brr}{\begin{array}}
\newcommand{\err}{\end{array}}
\newcommand{\bc}{\begin{center}}
\newcommand{\ec}{\end{center}}
\newcommand{\sss}{\scriptscriptstyle}
\newcommand{\bea}{\begin{eqnarray}}
\newcommand{\eea}{\end{eqnarray}}
\newcommand{\bean}{\begin{eqnarray*}}
\newcommand{\eean}{\end{eqnarray*}}

\section{Introduction.}
In this letter we analyse the infrared stability of $N=4$ supersymmetric gauge
theories. We consider the embedding of the supersymmetric
theory in a space of more general models which have the same field content
and gauge group as the supersymmetric one, but do not possess any
supersymmetry.
The symmetric theory is a fixed point of the renormalization group
equations. We discuss the infrared stability of such a solution.\par
It has been a long standing idea that the renormalization group flow could lead
to a more symmetric phase in the infrared (or in the ultraviolet).
This idea has been applied to several models~\cite{N1,W1,I1,C1,N2,Deg},
among which supersymmetric gauge model~\cite{C2,I2}.
In cases where the models considered have a residual $N=1$ supersymmetry
more powerful methods exist which rely on the holomorphicity properties of
the models and the structure of the exact beta functions~\cite{Shif,Leigh}.
This approach is also related to problems like the analysis of the stability
of BRS gauge invariant theory~\cite{Kraus} or the treatment of QCD with
boundaries~\cite{W2}.
Moreover, there are recent examples~\cite{S} of gauge theories in different
dimensions
which are conjectured to flow in the IR to non-trivial fixed points with
enhanced global symmetries.

In this letter we limit ourselves to the perturbative analysis of the
RG flow in a class
of gauge theories which represents a neighborhood of $N=4$ supersymmetric
gauge theory. All these models admit a lagrangian description and we consider
only
renormalizable theories (for an analysis of
infrared
stability of supersymmetric theories in terms of non renormalizable
effective lagrangians, see~\cite{Clark}).\par

We concentrate our attention on the stability of $N=4$ super Yang-Mills with
gauge group $SU(2)$.
This model has already been studied \cite{I2} in the case of a residual $N=1$
supersymmetry and turned out to be IR stable, while for groups others than
the so called safe algebras the same model gave an unstable $N=4$ fixed point.
We show that the stability in the case of $SU(2)$ can  be
ascribed to the residual supersymmetry of the model. We find indeed that
if we do not impose an $N=1$ residual supersymmetry, the $N=4$
supersymmetric theory becomes un IR unstable fixed point even in the case of
safe algebras.\par

In section 2 we briefly review the results for the $N=1$
supersymmetric case. In section 3 we extend this analysis to the case of a
general non supersymmetric model and we show that the $N=4$ fixed point is
unstable.

\section{$N=1$ supersymmetric case.}

$N=4$ supersymmetric gauge theories can be thought of as $N=1$ gauge theories
with three chiral superfields in the adjoint representation of the gauge group
$G$, $\Phi^i=\sum_a\Phi^i_aT^a$ ($i=1,2,3$;$a=1, \ldots ,dimG$),
coupled through the superpotential
$\lambda\varepsilon_{ijk}Tr(\Phi^i[\Phi^j,\Phi^k])$.
The group generators $T_a$ are  in the fundamental representation.
$N=4$ supersymmetry forces the coupling $\lambda$ of the chiral
superpotential to be equal to the gauge coupling $g$.\par
One can relax the $N=4$ constraint and consider a more general family of
models: they have the same field content as $N=4$ Super Yang-Mills (three
chiral superfields and one vector superfield), are invariant under the gauge
group $G$ and still realize only one of the four supersymmetries.\par
The generic theory belonging to this family is described by the following
 Lagrangian
\bea
{\cal L}&=&\int d^4\theta\sum_{i=1}^3Tr(e^{-gV}\bar{\Phi}_ie^{gV}\Phi^i)
+\frac{1}{64g^2}\int d^{2}\theta
Tr(W^{\alpha}W_{\alpha})+\nonumber\\
        &+&(\frac{i\lambda}{3!}\int d^2\theta\varepsilon_{ijk}
Tr(\Phi^i[\Phi^j,\Phi^k])+\frac{f_{ijk}}{3!}
\int d^{2}\theta Tr(\Phi^i\lbrace\Phi^j,\Phi^k\rbrace))+h.c.)+\nonumber\\
        &+&{\cal L}_{G.F}+{\cal L}_{F.P}\label{eqil}
\eea

\noindent where $f_{ijk}$ is a totally symmetric constant tensor,
$W_{\alpha}=\bar{D}^2(e^{-gV}D_{\alpha}e^{gV})$ is the gauge field strength,
and
${\cal L}_{G.F},{\cal L}_{F.P}$ are the usual gauge fixing and ghost superfield
Lagrangians~\cite{Bag,Sohn}.\par
The two terms $Tr(\Phi^i[\Phi^j,\Phi^k])$ and
$Tr(\Phi^i\lbrace\Phi^j,\Phi^k\rbrace)$ are the most general renormalizable
interactions compatible with $N=1$ supersymmetry.\par
The former term is the usual $N=4$ potential for the scalar
superfields, while the latter, forbidden in extended supersymmetry, is non
zero only for $G=SU(N\geq3)$ or $E_8$, i.e. for groups that admit
a totally symmetric invariant tensor.

$N=4$ supersymmetry~\cite{Brink} is recovered for the following values of the
parameters
$$\lambda=g\qquad f_{ijk}=0\qquad\forall i,j,k $$
and represents a line of fixed point of the renormalization group equations,
which, at one loop level, read ($a$ and $b$ are positive constants):
\bea
\beta_g&=&0\nonumber\\
\beta_{\lambda}&=&a\lambda(\lambda^2-g^2)\nonumber\\
\beta_{f_{ijk}}&=&bf_{ijk}(\lambda^2-g^2).
\eea

The behaviour of the beta functions in the parameter space around the line
of fixed points shows that their stability relies on the presence of the
coupling $f_{ijk}$ and is therefore gauge group dependent: for safe
algebras~\cite{Georgi},
which do not allow for the $f_{ijk}$ coupling, $N=4$ theory is an
infrared attractor, while it is not an attractor, neither infrared nor
ultraviolet, for other groups ($SU(N)$ or $E_8$).
Similar results can also be obtained, without any recourse to explicit one
loop computation, from simple considerations on the form of the beta
functions and the properties of $N=1$ supersymmetry~\cite{Leigh}.

\section{Nonsupersymmetric case.}

It is likely that the infrared stability of the fixed points line in the case
of safe algebras is simply a consequence of the $N=1$ residual supersymmetry.
This is indeed suggested by a similar analysis carried out in~\cite{C2}
on a class of gauge theories with the same field content as $N=2$ Super
Yang-Mills, which shows that the supersymmetric theory
is an infrared unstable fixed point.\par
In the case of interest here, i.e. $SU(2)$ gauge group, $N=1$ supersymmetry
forces the parameter space to be two dimensional: the gauge and the Yukawa
coupling
constants.

Without any supersymmetry constraint there are many more interactions allowed
among the fields of the model.
It is possible that some of these new couplings turns out to be relevant in the
neighborhood of the fixed point, introducing instability directions.

We consider the simplest renormalizable model one can build out of the same
fields as $N=4$
super Yang-Mills but without $N=1$ supersymmetry. We assume non abelian gauge
invariance to be an exact symmetry of our class of models; it could
eventually be spontaneously broken at one loop~\cite{Col,I2,C2}.\par

We choose to work in four component notation, where the $N=4$ multiplet is
represented by~\cite{Sohn}:
\begin{itemize}
\item $A_{\mu}$ is the gauge vector, with field strength
$F_{\mu\nu}=\partial_{\mu}A_{\nu}-\partial_{\nu}A_{\mu}+ie[A_{\mu},A_{\nu}],$
\item
$\lambda$ is a Majorana fermion which represents the gaugino,
\item
$\psi_i$,($i=1,2,3$), are three matter Majorana fermions,
\item
$A_i$,($i=1,2,3$), are three matter scalars,
\item
$B_i$ ($i=1,2,3$) are three matter pseudoscalars.
\end{itemize}
\noindent All fields are in the adjoint representation of
$SU(2)$.
\noindent With this choice of the fields, the on shell Lagrangian for this
models is
\bea
{\cal L}&=& Tr\lbrace
-\frac{1}{4}F_{\mu\nu}F^{\mu\nu}+\frac{i}{2}\bar{\lambda}D\lambda
+\frac{1}{2}(D_{\mu}A_i)^2+\frac{1}{2}(D_{\mu}B_i)^2
-ie(\bar{\psi}_i[\lambda,A_i]+\bar{\psi}_i\gamma_5[\lambda,B_i])+\nonumber\\
        &-&\frac{i\lambda}{2}\varepsilon_{ijk}(\bar{\psi}_i[\psi_j,A_k]
-\bar{\psi}_i\gamma_5[\psi_j,B_k])\rbrace-f_1[Tr(A_iA_i+B_iB_i)]^2+\nonumber\\
&+&2(f_1-f_2)[Tr(A_iA_i)Tr(B_iB_i)-(Tr(A_iB_i))^2]+f_3[Tr(A_iA_j+B_iB_j)]^2+
\nonumber\\
&+&2(f_5-f_3)[Tr(A_iA_j)Tr(B_iB_j)-Tr(A_iB_j)Tr(A_jB_i)]+\nonumber\\
&+&2f_4[(Tr(A_iB_j))^2-Tr(A_iB_j)Tr(A_jB_i)]+{\cal L}_{G.F}+{\cal
L}_{F.P}\label{eqlag}
\eea

\noindent where the covariant derivative is defined as
$D_{\mu}=\partial_{\mu}+ie[A_{\mu},]$
and the gauge fixing and ghost Lagrangians are the usual ones~\cite{IZ}.

Radiative corrections and one loop renormalizability require the introduction
of the additional gauge invariant interactions between
scalars and
pseudoscalars. In facts, these are all the possible
independent interaction terms renormalizable by power counting that one can
construct out of the scalars and the pseudoscalars of the model.\par
In principle, we could have considered a more general Lagrangian
with an independent coupling constant for each interaction term, getting a
larger parameter space.\par
The particular choice of the coupling constants of eq.~(\ref{eqlag}) is a
consequence of an additional $U(1)$ symmetry we imposed on the Lagrangian in
order to get the simplest possible model which exhibits an unstable
behaviour.\par
This is a non anomalous $R$-symmetry of the original $N=4$ Lagrangian, which
leaves the vector field unchanged and acts on the other fields as
\bea
A+iB&\rightarrow&e^{-i\theta}(A+iB)\nonumber\\
\psi_i&\rightarrow&e^{-\frac{1}{2}\gamma_5\theta}\psi_i\nonumber\\
\lambda&\rightarrow&e^{\frac{3}{2}\gamma_5\theta}\lambda\nonumber\\
A_{\mu}&\rightarrow&A_{\mu}.
\eea

The $U(1)$ symmetry makes the scalars and pseudoscalars
interactions symmetrical and so it reduces the parameter space of the theory:
the Yukawa terms for $A$ e $B$ have the same coupling constant and the four
scalars (pseudoscalars) couplings are related to the mixed terms $A^2B^2$.\par

Lagrangian~(\ref{eqlag}) reduces to the usual $N=4$ on-shell
Lagrangian~\cite{Brink} when
$$f_5=0,g=e=\lambda,f_i=g^2\qquad i=1,\ldots ,4,$$\
while we recover the $N=1$ models of Antoniadis et
al.(eq.~(\ref{eqil}))~\cite{I2} for
$$f_5=\lambda^2-g^2,e=\lambda,f_i=\lambda^2\qquad i=1,\ldots ,4.$$\par

To determine the fixed points of the model, we need to find the zeros of the
beta functions for the various couplings. We calculate them at one loop in
the minimal subtraction scheme with dimensional
regularisation.

At one-loop level the presence of the new quartic terms in the Lagrangian
does not affect the propagators and the $1PI$ functions which enter the
computation of the beta-functions for the Yukawa coupling constant $\lambda$.
So we expect to find the same $\beta_{\lambda}$ as in the $N=1$ case.
Similarly, since the one-loop beta function for the gauge coupling depends
only on the field content of the model, which is kept the same as in
$N=4$, we reproduce the $\beta_g=0$ result of $N=4$
case.

In the analysis of the infrared behaviour of this class of models, we are
interested in the relative evolution of the various coupling parameters with
respect to the scale independent gauge coupling $g$ ($\beta_g=0$). To this
purpose it is more convenient to define effective couplings ratios~\cite{C2}:
\be
E(t)=\frac{e^2(t)}{g^2(t)},\;\;\Lambda(t)=\frac{\lambda^2(t)}{g^2(t)},
\;\; F_{i}(t)=\frac{f_i(t)}{g^2(t)}\;\; i=1,\ldots ,5
\ee

\noindent In term of these new couplings, the one-loop beta-functions
considerably simplify and become:
\bea
\beta_E&=&24hg^2E(E-1)\nonumber\\
\beta_{\Lambda}&=&24hg^2\Lambda(\Lambda-1)\nonumber\\
\beta_{F_1}&=&hg^2(-24F_1+16F_1\Lambda+16EF_1+6-8\Lambda^2-8E^2-16E\Lambda
+\nonumber\\
& &+36F_1^2+16F_2^2+2F_4^2+2F_5^2-8F_2F_5+8F_4F_5-28F_1F_3+\nonumber\\
& &-4F_3F_4-4F_2F_3+8F_3^2-4F_1F_5-4F_3F_5-8F_2F_4)\nonumber\\
\beta_{F_2}&=&hg^2(-24F_2+16F_2\Lambda+16EF_2+6-8\Lambda^2-8E^2-16E\Lambda
+\nonumber\\
& &+36F_1F_2+10F_2^2-4F_1F_3-28F_2F_3-12F_1F_5+4F_3F_5+\nonumber\\
& &+2F_5^2+2F_4^2-8F_1F_4+4F_3F_4+6F_1^2)\nonumber\\
\beta_{F_3}&=&hg^2(-24F_3+16F_3\Lambda+16EF_3-6+8\Lambda^2+8E^2+\nonumber\\
& &-16E\Lambda+28F_1F_3-4F_2F_3+8F_3F_5-20F_3^2+\nonumber\\
& &+8F_4F_5-2F_5^2-2F_4^2-4F_3F_4)\nonumber\\
\beta_{F_4}&=&hg^2(-24F_4+16F_4\Lambda+16EF_4-6+8\Lambda^2+8E^2+\nonumber\\
& &-16E\Lambda+8F_1F_3+20F_2F_4-8F_2F_3-10F_3^2+12F_3F_5+\nonumber\\
& &-2F_5^2-16F_4^2+4F_1F_4-8F_4F_5)\nonumber\\
\beta_{F_5}&=&hg^2(-24F_5+16F_5\Lambda+16EF_5+8\Lambda^2+8E^2+\nonumber\\
& &-16E\Lambda+8F_1F_3+4F_1F_5-8F_2F_3+20F_2F_5-10F_3^2+\nonumber\\
& &-16F_5^2-2F_4^2-8F_4F_5+12F_3F_4)\label{beta}
\eea
\noindent where $h=16\pi^2$.

One can easely check that the point corresponding to the $N=4$ "phase"
$$F_5=0,E=\Lambda=F_i=1\qquad i=1,\ldots ,4,$$
is indeed a solution of system (\ref{beta}), while
a numerical analysis of the same  system shows the existence of other 25
fixed point
solutions in addition to the $N=4$ supersymmetric one.
They all correspond to ordinary gauge field theories with Yukawa
and scalar interactions, without any supersymmetry.\par
It is therefore excluded the possibility that the
class of models we considered could flow to a $N=1$
or $N=2$ supersymmetric theory.
The question is now whether the $N=4$ fixed point is stable or not.

The stability properties of a fixed point are determined by the
linearization of the renormalization group equations~(\ref{beta}),
that control the
behaviour of the couplings in the neighbourhood of the fixed point.  More
precisely, the fixed point is stable in the infrared if the matrix
representing the linearized system has all positive eigenvalues.\par
We computed eigenvalues and eigenvectors of the linearized system around each
fixed point. For simplicity we list here the main results only.\par
The $N=4$ fixed point represents a saddle point in the parameter space.
The linearized system around this point has indeed one negative eigenvalue:
\be
\rho_1=-3\qquad\rho_2=\rho_3=\rho_4=6\qquad \rho_{5,6}=3\pm\sqrt{3}\qquad
\rho_7=16.
\ee
The corresponding eigenvector represents an instability direction in the
parameter space. Moreover we find that, moving to the infrared along this
instability direction
the system is attracted toward a non supersymmetric fixed point, characterized
by the following values of the
couplings:
\be
E=\Lambda=1,\qquad F_1=F_2=0.757,\qquad F_3=F_4=0.352,\qquad F_5=0,
\ee

This point turns out to be the unique IR stable fixed point.
All others points are neither infrared nor ultraviolet attractors.\par

Combining this result with the previous ones~\cite{C2,I2}, we draw the
conclusion that supersymmetric gauge theory are always unstable in the
infrared, though attractive, for every choice of the gauge group.

\vskip 1.0truecm
I would like to thank L. Girardello for suggesting the subject and
for useful conversations. I would like to acknowledge helpful
discussions with G. Bottazzi, A. Pasquinucci, G. Salam and A. Zaffaroni.
This work was supported by the European Commission TMR programme
ERBFMRX-CT96-0045, in which M.P. is associated to the University of Torino.

\end{document}